\documentclass[aps,pre,
twocolumn,
groupedaddress,a4paper, amsmath, showpacs,
]{revtex4}
\usepackage{dcolumn}

\begin{document}

\preprint{LTU 2003-8}

\title{Stations, trains and small-world networks}

\author{Katherine A. Seaton}
\email{k.seaton@latrobe.edu.au}
\author{Lisa  M. Hackett}
\affiliation{Mathematics Department, La Trobe University, Victoria 3086,
Australia}

\date{November 11, 2003}

\begin{abstract}
The clustering coefficient, path length and average vertex degree of two urban train
line networks have been calculated. The results are compared with theoretical
predictions for
appropriate random bipartite graphs. They have also been compared with one another to
investigate the effect of architecture on the small-world properties.
\end{abstract}


\maketitle


\section{Introduction}{\label{intro}}
Since model small-world networks were first proposed some five years ago
\cite{Watts,Watts99}, to interpolate between the properties of regular and
random graphs, many investigations have been carried out which have
confirmed that the small-world phenomenon occurs in a variety of
real-world settings.

There are two common features which make apparently very different
networks all ``small-world'' \cite{Watts}. First,  the  \textit{characteristic path
length} between vertices is short, compared to what might be expected
based on the total number of vertices. This is the phenomenon which has
entered popular culture as ``six degrees of separation''. The second
feature, the one we are noting when we agree with a new acquaintance that
the world is indeed small, is that two vertices both linked to a third
vertex are likely to have a direct link also. This feature has been called \textit{clustering}, and while random graphs have the first
feature, they do not have the  second. Comparisons of real-world networks with random graphs
having the same number of vertices and edges indicate much higher
clustering occurs in real, small worlds. In a recent review \cite{Newman2} these real-world networks are roughly
grouped as social, biological, technological and information networks.

Among the first social networks to be studied in this context are the
Hollywood network of actors \cite{Watts, Watts99}, linked by having appeared in the same movie,
and the colloboration network of scientists \cite{Newman1}, linked by having published
together. By considering not just the people involved but also the media
that created their links, the network can be seen to have a \textit{bipartite}
structure \cite{bipart}, because nodes of one kind (actors/authors) can only
be directly linked to nodes of the other kind (movies/papers). In contrast
to the usual random
 graph, unipartite or one-mode projections of random bipartite graphs onto
vertices of the one kind inherit some amount of in-built clustering, from the
links to  common vertices of the other type \cite{bipart}.
It is thus more appropriate to compare the path length and clustering of such a real world network to the
averaged properties of random bipartite graphs with
the same number of vertices of each kind and the same probability distribution of links from one to the other type \cite{bipart,
marvel}. 

The study \cite{LMH02} reported here was motivated by two quite different
investigations into small-world properties of railway networks, one type of technological distribution
network. The first study (of the Boston subway network) \cite{boston}
takes as its definition of an edge that a physical piece of track should
connect two vertices (stations). This presents immediate difficulties
at the end of a line, where the concept of a local clustering
coefficient becomes moot and the definition (see (\ref{clust}) below) undefined. (To resolve this problem, the authors propose
that different properties should be studied; rather than clustering
coefficient and average or greatest path length, they define local and
global \textit{efficiency}, and \textit{cost} \cite{eff, boston}. Other authors have
chosen not to include such vertices in their calculation.) 

In the second study \cite{IRR}, of part of the railway network of India,
two stations are defined to be linked if it is possible to  embark at the first  and reach the second without having to change
trains, regardless of the number of intermediate stations, i.e., they lie on
the same line or route. This definition removes the ``end of the line''
problem noted above. However, it imposes the same kind of  bipartite
structure
\cite{bipart} as is present in the original study of the Hollywood
actors network \cite{Watts99, Watts}, and studies of
the collaboration network of scientists \cite{Newman1,bipart}
and of Fortune 1000 directors (sit together on the board of a
company) \cite{bipart, corporate}. However, this bipartite structure was not
referred to in the analysis provided in \cite{IRR}, unlike the somewhat
more whimsical study of the (artificial) Marvel comic book universe
\cite{marvel}, our third motivation. 

Therefore, we wanted to investigate
railway networks from this bipartite viewpoint, comparing the properties of
networks having different architecture with each other, and with those of
the appropriate random bipartite model. The two networks we have chosen are Boston \footnote{The
map of the Boston subway network was taken  October 4, 2002 from http://www.mbta.com/ .}, because of the earlier study, and Vienna
\footnote{The map was taken from http://www.wienerlinien/at/ .} because of different features in its layout. In the following section we
give the definitions of the network properties we calculate. In Section \ref{study}
we explain features of the systems that we studied and give our results. The paper
closes with discussion and a some conclusions.

\section{Network properties}
\subsection{Definitions}\label{defin}
There are two different definitions of the measure of cliquishness or
clustering in a network \cite{eco, Newman2} of $N$ nodes, which attempt to quantify the same
property (but which have not always been clearly distinguished). They both take values in the interval $[0,1]$.
The first definition \cite{Newman3}, which is convenient for analytic calculations, corresponds to what sociologists refer to as
transitivity, and  will be denoted by $C^{(\text{T})}$:

\begin{equation}
C^{(\text{T})}=\frac{3 \times \text{number of triangles in the network}}
{\text{number of connected triples of vertices}} . \label{trans}
\end{equation}

The second  definition is easier to apply
to an actual network, i.e., numerically.  Chronologically, this definition preceded (\ref{trans})
in the small-world literature \cite{Watts}. 
If node $v$ has $k_v$ neighbors, there are ${k_v}\choose{2}$ possible
connections between them. The local clustering coefficient of node $v$ is
the fraction of these that are actually present:
\begin{equation}
C_v=\frac{2\times\text{number of links between
neighbors of $v$}}{k_v(k_v-1)}\label{local}
\end{equation}
and then the clustering coefficient of the whole network is the
average of the $C_v$:
\begin{equation}
C=\frac{1}{N}\sum_v C_v. \label{clust}
\end{equation}

The number of other vertices to which a vertex $v$ is linked is its \textit{degree} $k_v$ and the average degree for
the network is $z$. If there are $K$ edges in the network, clearly 
\begin{equation}
K=\frac{Nz}{2}  \label{links}
\end{equation}
since each edge adjoins two of the vertices.

Except in limiting cases, it is probably not important to make too much of the differences between (\ref{trans}) and (\ref{clust});
after all, there are several different measures of network size in common usage. Define the \textit{distance} $d_{ij}$ between two nodes
$i,j$
to be the  length (in number of edges) of the shortest path between them. The \textit{diameter} $D$
of a network is then the greatest distance between any two randomly chosen nodes. The \textit{average path length}
$L$ is the average of the distances $d_{ij}$ between all pairs  of nodes:
\begin{equation}
L=\frac{\sum_{i \neq
j} d_{ij}}{N \cdot (N-1)} . \label{length}
\end{equation}
Analytically, it proves useful to define \cite{bipart} a \textit{typical path length} $\ell$ which is determined by finding when the 
sum of the average number of first $z_1(=z)$, second $z_2$ , \dots , $z_{\ell}$ neighbors of a vertex reaches $N$:
\begin{equation}
1+\sum_{i=1}^{\ell} z_i =N. \label{ell}
\end{equation}

Despite their differences, these various measures of network size scale as $\log(N)$ for a random graph; since there
is no preferential attachment, the clustering coefficient of a random graph (by either measure) is of order
$
C_{\text{random}}=\text{O}(N^{-1})$ \cite{Newman2}. More precisely, it is a result of some twenty year's standing in random graph
theory that $\ell =\log(N)/\log(z)$, while the clustering coefficient was shown \cite{Watts99} to be
\begin{equation}
C_{\text{random}}=\frac{2K}{N(N-1)}. \label{rand}
\end{equation}

The conditions for a network to
be small-world are that the characteristic path length is comparable to that of a random graph (with the same $N$ and $z$) while it
displays significantly more clustering.

\subsection{Formulae for random bipartite networks}\label{random}

For a bipartite network, the number of nodes $N$ refers to the vertices of the one-mode projection (e.g., the actors or,
in this case, the stations). The number of nodes of the other type (train lines) will be denoted $M$, the average
number of train lines a station lies on is $\mu$ and the average number of stations per line is $\nu$ so that
\begin{equation}
N \mu=M\nu. \label{munu}
\end{equation}
 The number of links, $K$, refers to the unipartite graph.

Newman et al.~\cite{bipart} have given expressions for average values of the network properties defined above, for  various
distributions of vertex degree, in terms of their generating functions. In the case of
bipartite graphs, the appropriate expressions are summarized below.  

For specific real-world graphs, such as these two train networks, one knows the degree of each vertex, so that the generating functions
are polynomials. Let
$p_j$ be the probability that a station appears on $j$ train lines, and $q_k$ be the probability
that a train line has $k$ stations on it. Then the two generating functions needed are:
\begin{equation}
f_0(x)=\sum_j p_jx^j, \qquad g_0(x)=\sum_k q_kx^k. \label{generate}
\end{equation}
Defining further generating functions in terms of these (their interpretation is given in \cite{bipart} but
is  not needed  for calculations)
\begin{equation}
f_1(x)=\frac{1}{\mu}f_0'(x), \qquad g_1(x)=\frac{1}{\nu}g_0'(x),
\end{equation}
then the number of first and second neighbors of a station chosen at random are
\begin{subequations}\label{neigh}
\begin{align}
z_1&=f'_0(1)g_1'(1),\label{neigh1}\\
z_2&=f'_0(1)f_1'(1)[g_1'(1)]^2\label{neigh2};
\end{align}
\end{subequations}
the typical path length (\ref{ell}) is
\begin{equation}
\ell=\frac{ \log [(N-1)(z_2-z_1)+z_1^2]- \log z_1^2}{ \log (z_2/z_1)};
\end{equation}
and the clustering coefficient (\ref{clust}) is
\begin{equation}
C=\frac{M}{N}\frac{g_0'''(1)}{G_0''(1)}. \label{biclust}
\end{equation}
Here $G_0(x)$ has interpretation as the generating function of the probability distribution of number of first neighbors (i.e., degree)
on the unipartite graph of stations, and is defined by
\begin{equation}
G_0(x)=f_0(g_1(x)).
\end{equation}

Of course, the role of the nodes could be interchanged, i.e., a projection made onto
a one-mode graph of train lines rather than stations. In the case of company directors, studying the \textit{interlock} of
Fortune 1000 companies (rather than degree distribution for the network of directors) has provided explanation for the adoption of common
corporate policies and practices
\cite{corporate}.

\section{The small worlds of Boston and Vienna}
\subsection{This study: stations and trains}\label{study}

In early sociological studies of networks, getting a good definition of connection was sometimes problematic \cite{Watts99}. While
for technological networks this is generally less of a problem,  recall
the definition of a link between two stations used in this study: that one can travel between them without changing train. The
railway networks were treated as undirected and unweighted; if more than one train line can be used to travel from a particular station
to a second, with no change of train, in either direction, this counted as a connection. Further,  no weighting was made for
timetable information, such as the number of trains running on a line per day or whether trains ever ``run express''.

The subway network of Boston consists of four lines, serving 124 stations \footnote{Note
that the new silver line was incomplete at the time of \cite{boston}, so
we did not include it in our study either.}. The network is decentralized, in that no station lies on
all four lines, thus having every other station as a direct neighbor with our definition of adjacency. In fact, the highest value of
$k_v$ is 97. The diameter of the network is $D=3$.
One line splits into two branches and another into four. Consistent with our definition of connectivity, these were treated  as
separate lines, giving eight in total, some of which have a series of stations in common. 

There are five lines serving 76 stations in the U-Bahn network of Vienna, which is also decentralized. These lines interact with each
other at single stations, though possibly more than once due to their tendency to follow curved (rather than radial) paths.
The highest degree is $k_v=43$ and the diameter is also 3.

From the network maps, the  value of $d_{ij}$ for each pair of stations was
entered into a spreadsheet. Using various spreadsheet functions, the degree of
each station $k_v$, the total number of links $K$, the average degree $z=z_1$, the clustering
coefficient
$C$ as defined in (\ref{clust}), and the average path length $L$ defined in (\ref{length}) were calculated \cite{LMH02}. 
By counting the instances when $d_{ij}=2$, $3$, the average number of second $z_2$ and third $z_3$ neighbors could also be
determined. These `Actual'  values appear in Table \ref{calculations}. 

\begin{table*}
\begin{ruledtabular}
\begin{tabular}
{lcccc} 
 &  \multicolumn{2}{c}{Boston} &
 \multicolumn{2}{c}{Vienna} \\
 &  Actual & Theory  & Actual & Theory \\
\hline Number of stations, $N$ &  \multicolumn{2}{c}{124} & \multicolumn{2}{c}{76} \\
Number of train lines, $M$  & \multicolumn{2}{c}{8} & \multicolumn{2}{c}{5} \\
Number of links, $K$  & {1711} &1936& {785}&788 \\
Average degree, $z_1=z$ &  27.597 & 31.226
& 20.658 & 20.737 \\
Mean 2nd neighbors, $z_2$ &  91.048 & 766.75 &
43.842 & 97.213 \\
Mean 3rd neighbors, $z_3$&4.355&&10.500&\\
Network size, $L$, $\ell$ &  1.8110 & 1.4187 & 1.8646 & 1.7236 \\
Clustering coefficient, $C$, $C^{(\text{T})}$ &  0.9276 &
0.4808 & 0.9450 & 0.7989 \\
$C_{\text{random}}$ &  & 0.2244 & & 0.2754  \\
\end{tabular}
\end{ruledtabular}
\caption{Network properties for each train network. The results in the `Actual' columns were obtained from
the network maps (by spreadsheet calculations). Apart from $C_{\text{random}}$, the results in the `Theory' columns were obtained from
the formulae in Section \ref{random} for random bipartite graphs with appropriate specified probability distributions, using
\textsc{Maple}.}
\label{calculations}
\end{table*}
Also from the maps, the discrete probability distributions $p_j$ and $q_k$ required for using the generating 
functions (\ref{generate}) were identified. These appear in Tables \ref{p} and \ref{q}. The mean properties of 
random bipartite graphs with the same distributions of links from one type of node to another, $z_1$, $z_2$, $\ell$ and $C^{\text{(T)}}$,
were calculated using equations (\ref{neigh})--(\ref{biclust}), and appear in Table \ref{calculations} in the `Theory' columns, together
with the total number of (predicted) links, calculated using $z_1$ in (\ref{links}). Finally, using (\ref{rand}), $C_{\text{random}}$ was
calculated for the ``usual'' random graph.
\begin{table}
\begin{ruledtabular}
\begin{tabular}{rcccccc}
   $j:$ &   1& 2&3&4&5&6\\
\hline Boston & 100&12&3&5&3&1\\
Vienna & 67 & 8&1& & &\\
\end{tabular}
\end{ruledtabular}
\caption{Number of stations lying on $j$ lines, i.e., $N p_j$. }
\label{p}
\end{table}

\begin{table}
\begin{ruledtabular}
\begin{tabular}{rccccccccc}
$k:$&7&12&14&18&19&20&21&24&33\\
\hline
Boston&&1&&1&1
&1&&3&1\\
Vienna&1& &1& & &1&1&1& \\
\end{tabular}
\end{ruledtabular}
\caption{Number of lines containing  $k$ stations, i.e. $Mq_k$. }
\label{q}
\end{table}

Having $f_0(x)$ and $g_0(x)$ exactly, \textsc{Maple} was used to find the generating
function for the probability distribution $r_z$ of number of stations $z$ a randomly chosen station is linked to:
\begin{equation}
G_0(x)=f_0(g_1(x))=\sum_z r_z x^z. \label{costat}
\end{equation}
By swapping the roles of $f$ and $g$ the probability distribution for interactions (i.e., interlock) between
train lines could also be determined. 

\subsection{Discussion}

From Table \ref{calculations}, both networks satisfy the two basic conditions for being small-world. The size of the 
networks, measured by diameter ($D=3$ for both) or by average path length $L$, is small when compared to the number of
vertices. The actual clustering coefficients are much greater than $C_{\text{random}}$ for appropriate $N$, $K$.
While  in relative terms the
ratio $C/C_{\text{random}}$ is orders of magnitude smaller than, say, that observed for the Hollywood network \cite{Watts, bipart}, it is
comparable with that observed in studies of networks of similar size as listed in Table \ref{sample}. 

\begin{table}
\begin{ruledtabular}
\begin{tabular}{ccccc}
Network&$N$&$K$&$C/C_{\text{random}}$&Reference\\
\hline \hline
Boston
&124&1711&4.13&\\
Vienna&76&785&3.43&\\
\hline
Dolphins&64&159&3.74&\cite{dolphin}\\
China airports&128&1165&5.13&\cite{ANC}\\
\textit{C. elegans}&282&1974&5.60&\cite{Watts, Watts99}\\
Hollywood&450,000&2.55$\times 10^7$&789&\cite{bipart}\\
\end{tabular}
\end{ruledtabular}
\caption{Actual clustering coefficients compared to $C_{\text{random}}$ for appropriate network size $N$ and number of links $K$.}
\label{sample}
\end{table}

However, in absolute terms, for both networks $C$ is very much  higher than the values reported in other studies.
Our definition of connectivity has resulted in $C$ being so close to 1 
because of the proportion of stations in each network ($ \gtrsim$ 80\%, see Table \ref{p}) which lie on a single train line. 
All
stations to which they are connected are also linked to one another, forming a clique, and each giving $C_v=1$ as their contribution to
(\ref{clust}). The most useful comparison may be to the railway network of India \cite{IRR} which has  $C=0.69$. A country-wide train
network is not constrained as an urban subway network is, and train routes criss-cross between stations in different patterns, giving
lower local clustering.

However, some amount of this clustering is expected solely on the basis of the underlying bipartite structure \cite{bipart},
and the calculation of $C^{\text{(T)}}$ gives a measure of it. Although they are not precisely the same quantity, comparing
the actual value of $C$ to the theoretical $C^{\text{(T)}}$ for the same distributions $p_j$ and $q_k$ shows that the Boston network 
appear to have more genuine or ``excess'' clustering. For comparision, in Table \ref{comp} some other values from the literature are
listed. In social networks, this clustering is explained by people introducing their friends or collaborators to one another
\cite{Watts, bipart} while for the artificial world of Marvel comics it arises from the way that the (6 486) characters are not
distributed randomly across all the (12 942) comic books but appear in teams, in series \cite{marvel}. 
The Vienna network has a structure similar to that originally proposed to model small-worlds \cite{Watts, Watts99}: it consists
of cliques of connected nodes (stations on a single line) which interact with one another through certain nodes (stations common to two
lines) which connect the network but have very low values of $C_v$. On the other hand, because of the branched lines in the Boston
network, whole groups of stations are shared between lines, overlapping the cliques and enhancing $C_v$ for the shared vertices.
\begin{table}
\begin{ruledtabular}
\begin{tabular}{ccccc}
Network&$C$ (actual) &$C^{\text{(T)}}$ (theory)&$C/C^{\text{(T)}}$&Ref.\\
\hline \hline
Boston
&0.9276&0.4804&1.93&\\
Vienna&0.9450&0.7989&1.18&\\
\hline
Fortune 1000&0.588&0.590&1.00&\cite{bipart}\\
Physics co-authors &0.452&0.019&2.35&\cite{bipart}\\
Hollywood &0.199&0.084&2.36&\cite{bipart}\\
Marvel &0.192 &0.0066 &29.09&\cite{marvel} \\
\end{tabular}
\end{ruledtabular}
\caption{Comparison of actual and theoretical clustering for a number of one-mode networks with underlying bipartite structure.}
\label{comp}
\end{table}

The average degree $z$ as predicted by the random bipartite theory agrees well with the actual observed values (see Table
\ref{calculations}). Using $z$ to calculate the total number of links $K$ shows that this agreement is better for Vienna
than Boston, where the prediction is for an extra 200-odd links between the 124 stations. This is also attributable to
the effect of the branched lines, causing stations to interact repeatedly with groups of common stations, rather than having
links to as many new stations as might be expected based on the number of lines on which they lie.

The worst discrepancy between the predictions of the random bipartite model and the actual results was in estimating the
number of second neighbors $z_2$. In both cases, the predicted number exceeded the actual average
number of second neighbors. This is a flow-on from  the high local clustering; if a node and most of its
neighbors form a clique, there are few paths of length two.
But the theoretical $z_2$ also exceeded the total number
of nodes in the network, in the case of Boston by an order of magnitude!

There are several limitations to our study which may have affected how the results compare with the bipartite theory. The small-world
effect is notable in systems which are not only connected and decentralized (as ours are) but also \textit{large}, and \textit {sparce}
\cite{Watts}. Though the value of $N$ is not extremely large, it is comparable to the  other studies cited in Table
\ref{sample}. Sparceness refers to the number of links $K$ in comparison to the number of nodes $N$, specifically that
\begin{equation}
K\ll {N\choose 2}\qquad \text{or} \qquad z \ll N-1.
\end{equation}
We have $z/(N-1) \approx 0.25$ which is not particularly small, and is certainly larger than the studies cited in Table \ref{sample}.
One feature of our study which appears unique is that $M/N \approx 1/15$. For other networks which have been studied from
the bipartite point of view this ratio takes values: 0.5 (Marvel universe \cite{marvel}), 1.0 (Indian railroad \cite{IRR}), 2.0
(Hollywood
\cite{Watts99, Newman2}), and 8.9 (Fortune 1000 \cite{corporate, Newman2}).
We speculate that the balance of the various sizes ($N$, $M$, $K$) may be the main problem with the second degree estimate, but since no
other authors have calculated or commented on
$z_2$ in their studies we cannot offer further support for this explanation.  (We hence do not give too much attention to
the calculated value of
$\ell$ since it  relies on both $N$ and $z_2$.)

Finally we comment on the distributions of linked stations and interlocked train lines, though we do not give the corresponding long
but rather sparce tables of actual values. The distribution $r_z$ from (\ref{costat}) for both Vienna and Boston corresponded closely to
the actual values, with some smoothing and a long, low tail as would be expected. The analogous generating function, for how many lines a
randomly-chosen line shares stations with, gave reasonable agreement with the actual values for Vienna. However, the
expression for Boston suffered from a similar problem to the $z_2$ results referred to above; it was a very smeared-out distribution
which gave the highest probability to 22 lines, though actually the maximum possible is
$M-1=7$. Presumably this problem is due not so much to $M$ being small (since Vienna would have had that problem too), but mainly to
the unusual actual distributions which result from the common stretches of line, which the averaged distribution from the random model
struggles to mimic. In this context, it should be noted that the theoretical interlock of company directors \cite{bipart} was not in
close agreement with the actual data, and the proferred explanation is similar (when translated into more human terms).

\subsection{Conclusion}

To quote the paper \cite{bipart} that introduced analysis based on bipartite structure to the small-world literature: \textit{``[It] is
perhaps best to regard our [bipartite] random graph as a null model---a baseline from which our expectations about network structure
should be measured. It is deviation from the random graph behavior, not agreement with it, that allows us to draw conclusions about
real-world networks.''} Although there have been other studies of small-world properties of transport networks \cite{ANC}, including
railway networks
\cite{boston}, this study appears to be the first that has not only had underlying bipartite structure \cite{IRR}, but has used
this in its analysis.  Further, because we have
studied two networks at once, we have been able to draw comparisons between them based on their different train-line architecture.
We have seen properties in common, such as high $C$, properties close to prediction, e.g., $z$, and properties which vary from the random
bipartite graph model, and we have attempted to explain the behavior in terms of our definition of connectivity and how this works
itself out in the actual networks.  This study extends the application of bipartite analysis from social affiliation networks
\cite{bipart, marvel} to the technological; presumably there are examples to be found also among real-world information and biological
networks.

\begin{acknowledgments}
We thank the authors of \cite{marvel} for sending us a revised version of
their manuscript, and Jerry Davis for helpful correspondence.
\end{acknowledgments}

\end{document}